\documentstyle[12pt]{article}
\newcommand{\QED}{\mbox{\rule[-1.5pt]{6pt}{10pt}}}
\newcommand{\Ran}{{\rm Ran\,}}

\newcommand{\dist}{{\rm dist\,}}

\newcommand{\DD}{{\cal D}}

\newcommand{\HH}{{\cal H}}

\newcommand{\KK}{{\cal K}}

\newcommand{\PP}{{\cal P}}
\newcommand{\RR}{{\cal R}}
\newcommand{\SS}{{\cal S}}
\newcommand{\TT}{{\cal T}}

\newcommand{\nid} {\noindent}

\def\Ran{\mbox{Ran}\, }
\def\sgn{\mbox{sgn}}
\font\sr = msbm10 scaled \magstep 1
\font\indsr = msbm9

\newcommand{\C}{\mbox{\sr C}} 
\newcommand{\N}{\mbox{\sr N}} 
\newcommand{\R}{\mbox{\sr R}} 
\newcommand{\T}{\mbox{\sr T}} 
\newcommand{\Z}{\mbox{\sr Z}} 
\newcommand{\LB}{{\mbox{\sr L}}} 

\newcommand{\Nin}{\mbox{\indsr N}}

\newcommand{\Zin}{\mbox{\indsr Z}}

\def\pa {\partial}

\newcommand{\be}{\begin{equation}}
\newcommand{\ee}{\end{equation}}
\newcommand{\bea}{\begin{eqnarray}}
\newcommand{\eea}{\end{eqnarray}}

\newcommand{\al}{\alpha}
\newcommand{\La}{\Lambda}
\newcommand{\la}{\lambda}
\newcommand{\w}{\omega}
\newcommand{\bt}{\beta}
\newcommand{\sg}{\sigma}
\newcommand{\ga}{\gamma}
\newcommand{\Ga}{\Gamma}
\newcommand{\ro}{\rho}
\newcommand{\vp}{\varphi}
\newcommand{\dl}{\delta}
\newcommand{\DL}{\Delta}

\newcommand{\lb}{\la(\bt)}
\newcommand{\HK}{\hat K}
\newcommand{\HV}{\hat V}
\newcommand{\HC}{\hat C}
\newcommand{\GL}{\Gamma_\lambda}
\newcommand{\gbl}{g(\beta,\lambda)}
\newcommand{\Gbl}{G(\beta,\lambda)}
\newcommand{\Pbl}{\Phi(\beta,\lambda)}
\newcommand{\bo}{\bar 0}
\newcommand{\io}{\bar\iota}
\newcommand{\tps}{\tilde\psi}
\newcommand{\tvp}{\tilde\vp}
\newcommand{\vf}{\vp^f}
\newcommand{\vn}{\vp^{(N)}}
\newcommand{\tch}{\tilde\chi}
\newcommand{\FL}{F(n)-\la}
\newcommand{\PS}{P_{\SS}}
\newcommand{\QS}{Q_{\SS}}
\newcommand{\pr}{{\rm pr}}
\newcommand{\dn}{\dist(n_1,\pr_1(\SS\setminus\{n\}))}
\newcommand{\lw}{(2|\la|/\w)}
\newcommand{\lww}{\left({2|\la|\over\w}\right)}
\newcommand{\las}{\la_\star}
\newcommand{\BL}{B(\la)}
\newcommand{\bpl}{\beta_{\pm}(\lambda)}

\newcommand{\ct}{} 

\begin{document}


\font\fifteen=cmbx10 at 15pt
\font\twelve=cmbx10 at 12pt

\begin{titlepage}

\begin{center}

\renewcommand{\thefootnote}{\fnsymbol{footnote}}

{\twelve Centre de Physique Th\'eorique\footnote{
Unit\'e Propre de Recherche 7061
}, CNRS Luminy, Case 907}

{\twelve F-13288 Marseille -- Cedex 9}

\vspace{1 cm}

{\fifteen PERTURBATION OF AN EIGEN-VALUE FROM \\
A DENSE POINT SPECTRUM : AN EXAMPLE}

\vspace{0.3 cm}

\setcounter{footnote}{0}
\renewcommand{\thefootnote}{\arabic{footnote}}

{\bf 
P. Duclos\footnote{and
PhyMaT, Universit\' e de Toulon et du Var, BP 132,\\
\null\qquad F-83957 La Garde cedex, France\\
\null\qquad duclos@naxos.unice.fr},
P.\,\v{S}\v{t}ov\'{\i}\v{c}ek\footnote{Department of Mathematics, Faculty of
Nuclear Science,  CTU, Trojanova 13,\\
\null\qquad  CZ-120 00 Prague, Czech
Republic\\
\null\qquad stovicek@kmdec.fjfi.cvut.cz} and M.
Vittot\footnote{vittot@cpt.univ-mrs.fr}  }

\vspace{1 cm}

{\bf Abstract}

\end{center}

We study a perturbed Floquet Hamiltonian $K+\bt V$ depending on a
coupling constant $\bt$. The spectrum $\sigma(K)$ is assumed to be pure
point and dense. We pick up an eigen-value, namely $0\in\sigma(K)$, and
show the existence of a function $\la(\bt)$ defined on $I\subset\R$ such
that $\lb\in\sigma(K+\bt V)$ for all $\bt\in I$, 0 is a point of
density for the set $I$, and the Rayleigh-Schr\"odinger perturbation
series represents an asymptotic series for the function $\lb$. All ideas
are developed and demonstrated when treating an explicit example but
some of them are expected to have an essentially wider range of
application.

\vspace{1 cm}


\bigskip

\noindent February 1997

\noindent CPT-97/P.3454

\vspace{0.5 cm}

\noindent anonymous ftp or gopher: cpt.univ-mrs.fr

\renewcommand{\thefootnote}{\fnsymbol{footnote}}

\end{titlepage}

\setcounter{footnote}{0}
\renewcommand{\thefootnote}{\arabic{footnote}}

\tableofcontents
\vspace{1cm}

\section{Introduction}%

A common problem occurring frequently in theoretical physics is the
eigen-value problem for a perturbed operator $K+\bt V$, with $\bt$ being
a coupling constant, under the assumption that $F_0$ is a known
eigen-value of the unperturbed operator $K$. The Rayleigh-Schr\"odinger
(RS) series gives a formal solution $F(\bt)$, with $F(0)=F_0$, as an
unambiguously determined formal power series. The regular perturbation
theory due to \ct {Rellich (1937)} and \ct {Kato (1966)} justifies this
formal series as an analytic function well defined on a neighbourhood of
$\bt=0$ provided one essential condition is fulfilled -- the eigen-value
$F_0\in\sigma(K)$ must be {\em isolated}. On the other hand, the situation
when an eigen-value of $K$ is not isolated is far away of being
exceptional and recently attracted a considerable attention (see
\ct {Simon 1993} and references therein).

So called Floquet Hamiltonians represent a class of operators having
even a dense pure point spectrum in many interesting examples. They were
introduced as an important tool to study time-dependent systems (see
\ct {Howland 1979}, \ct {Yajima 1977}). A distinguished subclass is
formed by the systems with the potential $V(t)$ being $T$-periodic and
bounded. The period is usually considered as a parameter. After
rescaling the time, the potential $V(t)$ becomes $2\pi$-periodic and the
frequency $\omega=2\pi/T$ appears in front of the time derivative. Thus
one is lead to study the operator $K+\bt V(t)$ acting in
$\KK:=L^2(\T,dt)\otimes\HH$, with $\T=\R/2\pi\Z$, and
$$
K:=-i\w\pa_t+H,\quad \w>0\,,
$$
where $H$ is the "true" Hamiltonian acting as a self-adjoint operator in
a separable Hilbert space $\HH$. We use the loose notation identifying
$\pa_t$ with $\pa_t\otimes 1$, $H$ with $1\otimes H$ etc. Provided the
spectrum $\sg(H)$ is pure point the same is true for
$\sg(K)=\w\Z+\sg(H)$. It is known that $\sg(K)$ is dense in $\R$ for
almost all $\w>0$ as soon as $\sup \sg(H)=+\infty$. Recently the spectrum
of $K+\bt V(t)$ has been studied by the aid of a quantum version of the
KAM method due to \ct {Bellissard (1983)} (see also \ct {Combescure
1987}, \ct {Bellissard, Vittot 1990}, \ct {Bleher, Jauslin, Lebowitz 1992}, 
\ct  {Duclos, \v{S}\v{t}ov\'{\i}\v{c}ek 1996}) as well as by adiabatic tools
(\ct  {Nenciu 1993}, \ct {Joye 1994}).

In the present paper we focus on a particular example with
$\HH=L^2(\T,dx)$,
\be
\label{model}
H=-\pa_x^{\,2}\ (\mbox{+ periodic boundary conditions}),\quad
V(t)=4 \cos t \cos x \,.
\ee
Clearly, $\sg(H)=\{ E(k)=k^2\,;\ k\in\Z\}$ and so
$\sg(K)=\{ F(n)=\w n_1+E(n_2) \,;\ n\in\Z\times\Z\}$. The spectrum of
$H$ is degenerate and that makes the problem more complicated; the only
non-degenerate eigen-value is $E(0)=0$. This is why we restrict
ourselves to eigen-values $F(n)$ of $K$ with $n_2=0$. In order to be
specific, we shall even consider the only eigen-value $F(0)=0$. We are
going to address the question whether there exists an eigen-value $\lb$
of the operator $K+\bt V(t)$ which could be considered as a perturbation
of $F(0)=0$ depending on the parameter $\bt$. A possible answer is
given in
\proclaim{Proposition 1}.
For almost all $\w>0$, there exists a real-valued function $\lb$ defined
on $I\subset\R$ with the properties:\\
(1) for $\forall\beta\in I$,
$\lambda(\beta)$ is an eigen-value of $K+\beta V(t)$,\\
(2) $\lim_{\delta\downarrow 0}
|I\cap[-\delta,\delta\,]|/2\delta=1$,\\
(3) the function $\lambda(\beta)$ has an
asymptotic expansion at $\beta=0$ coinciding with the formal
Rayleigh-Schr\"odinger perturbation series for the eigen-value $F(0)=0$
of $K$.

In fact, our final goal (not achieved in this paper) is to prove a
similar proposition for a much wider class of Floquet Hamiltonians.
However, as this program seems to be extremely complex, we preferred to
develop and to demonstrate the main ideas when treating an explicit
example. But the proof, even in the case of our very particular model, is
far away of being obvious and straightforward. We note that the essential
assumptions which are expected to be required also in the general case
are a sufficient smoothness of $V(t)$ (generally the order of the
asymptotic series depends on the order of differentiability of $V(t)$)
and a gap condition imposed on the eigen-values of $H$:
$\sg(H)=\{E(k)\,;\ k\in\Z_+\}$ and
\be
\label{gap}
\inf_{k\in\Zin_+}
{E(k+1)-E(k)\over (k+1)^{\alpha}} =: C_E>0
\quad\mbox{ for some } \al>0 \,,
\ee
(basically $\al=1$ in our example when overlooking the degeneracy).
Apparently, our model captures already all basic features but, on the
other hand, it makes possible some simplifications and can be treated on
a relatively elementary level. The rest of the paper is devoted to the
proof of Proposition 1 but, whenever possible, we shall try to consider a
more general situation and to propose some ideas applicable also to
other models.

\section{Basic equation}%

The starting point is the eigen-value equation for $K+\bt V$. Assume that
0 is a non-degenerate eigen-value of $K$ and $f$ is the normalized
eigen-vector. Let $P$ be the orthogonal projector onto the eigen-space
$\C f$ and $Q:=1-P$. We are seeking $\la=\lb\in\R$ and $g\in\KK$ such
that $Pg=0$ and
\be
\label{start}
(K+\bt V)(f+g)=\la(f+g) \,.
\ee
Without loss of generality we can assume that
\be
\label{PVP}
PVP=0 \,.
\ee
Apply successively the projectors $P$ and $Q$ to the equation
(\ref{start}). The result is
\bea
\label{basicl}
\la &=& \bt\,\langle Vf,\, g\rangle \,, \\
\label{basic}
(\HK+\bt\HV-\la)g &=& -\bt QVf \,.
\eea
Here and everywhere in what follows the hat indicates the restriction to
$\Ran Q$ in the sense: $\hat X=QXQ|\, \Ran Q$.

According to our assumptions, $\HK$ is invertible and we set
$\Ga_0:=\HK^{-1}$ (defined on $\Ran Q$). For $\la\not\in\sg(\HK)$ we
define also
$$
\GL:=(\HK-\la)^{-1}=(1-\la\Ga_0)^{-1}\Ga_0 \,.
$$
Keeping $\la$ as an auxiliary parameter one can solve formally
(\ref{basic})
\be
\label{gbl}
g=\gbl:=-\bt(1+\bt\GL\HV)^{-1}\GL QVf \,.
\ee
Plugging (\ref{gbl}) into (\ref{basicl}) we get a fixed-point equation
for the eigen-value $\la=\lb$,
\bea
\label{fixed}
&& \la=G(\bt,\la)\quad\mbox{where} \cr
&& G(\bt,\la):=-\bt^2\langle QVf,\,(1+\bt\GL\HV)^{-1}
\GL QVf\rangle\,.
\eea

The trick with the projectors and keeping $\la$ as an auxiliary
parameter is well known and related to various names. In the regular
case, when $d:=\dist(0,\sg(\HK))>0$, one can rederive this way
Rellich-Kato Theorem. Indeed, we have $\|\Ga_0\|=d^{-1}$ and
$(1+\bt\GL\HV)$ is invertible (on $\Ran Q$) provided $|\bt|$ and
$|\la|$ are sufficiently small. The implicit function theorem applied
to (\ref{fixed}) then gives the result.

To solve (\ref{fixed}) formally one can use B\"urmann-Lagrange
Formula which can be proven with some combinatorics and not 
necessarily with the Cauchy Residuum Theorem. Write
\bea
\Gbl &=& \sum_{M=0}^{\infty}\Phi_M(\bt)\,\la^M,\quad\mbox{where}\cr
\Phi_M(\bt) &=& -\sum_{k=1}^{\infty}\,
\sum_{\mu\in\Nin^k,\ |\mu|=k+M}
(-\bt)^{k+1}\,\langle QVf,\, \HK^{-\mu_1}\HV\HK^{-\mu_2}\dots
\HV\HK^{-\mu_k}QVf\rangle \,. \nonumber
\eea
The formal solution $\lb$ reads
\be
\label{RS}
\lb=\sum_{N=1}^\infty \sum_{\nu\in\TT(N)}
\Phi_{\nu_1}(\bt)\dots\Phi_{\nu_N}(\bt) =
\sum_{M=2}^\infty \xi_M\,\bt^M \,,
\ee
where $\TT(N)\subset\Z_+^{\,N}$ is the set of rooted $N$-trees:
$\nu=(\nu_1,\dots,\nu_N)\in\TT(N)$ iff $\nu_k + \dots + \nu_N\le N-k$,
$2\le k\le N$, and $|\nu|=N-1$. Consequently, one gets an expression for the 
coefficients $\xi_M$
\bea
\label{xi1}
\xi_M &=& \sum_{N=1}^{[M/2]}\sum_{\nu\in\TT(N)}\
\sum_{k(1),\dots,k(N)\in\Nin}\
\sum_{\mu(1)_\in\Nin^{k(1)},\dots,\mu(N)_\in\Nin^{k(N)} } \cr
&& \times(-1)^{M+N}\prod_{j=1}^N
\langle QVf,\, \HK^{-\mu(j)_1}\HV\HK^{-\mu(j)_2}\dots
\HV\HK^{-\mu(j)_{k(j)} }QVf\rangle \,,
\eea
with the summation range being restricted by
$$
k(1)+\dots+k(N)+N=M,\quad\mbox{and }\ |\mu(j)|=k(j)+\nu_j,\
1\le j\le N\,.
$$

Of course, this result must coincide with the standard RS perturbation
series written in the form (see \ct {Kato 1966})
\be
\label{xi2}
\xi_M={(-1)^M\over M}\ \sum_{k_1+\dots+k_M=M-1,\ k_i\ge 0}
{\rm tr}\left(V\hat R^{k_1}\dots V\hat R^{k_M}\right) \,,
\ee
where the symbol $\hat R^{k}$ is defined by: $\hat R^{0}=-P$  , and for
$k\ge 1$, $\hat R^{k}|\Ran P=0$, $\hat R^{k}|\Ran Q=\HK^{-k}$. The
equality between (\ref{xi1}) and (\ref{xi2}) can be verified quite
straightforwardly using (\ref{PVP}) and the following fact:
\proclaim{Lemma 2}.
For a given $N\in\N$ and each $\sg=(\sg_1,\dots,\sg_N)\in\Z_+^{\,N}$
obeying $|\sg|=N-1$ there exists exactly one cyclic permutation of
$\sg$, $\sg'=(\sg_{N-m+1},\dots,\sg_N,\sg_1,\dots,\sg_{N-m})$
(determined by $m\in\{0,1,\dots,N-1\}$), such that $\sg'\in\TT(N)$.

Hence each term of (\ref{xi1}) is a grouping of many terms of (\ref{xi2})
where we take into account the cyclic property of the trace.

However in the case when $\sg(K)$ is dense in $\R$ and so $\dist(0,
\sg(\HK))=0$ it seems to be hopeless to consider the RS series as a
convergent series. The complication comes from arbitrarily large powers of
$\HK^{-1}$ in (\ref{xi1}) (or (\ref{xi2})) since among eigen-values of
$\HK$ there are arbitrarily small numbers -- so called small
denominators. Probably the maximum one can attempt in this situation is
to verify the finiteness of the coefficients $\xi_M$ (generally up to
some order depending on the smoothness of $V(t)$) and to show that the
RS series is asymptotic for the function $\lb$.

Let us specify the formula (\ref{xi1}) to our example (\ref{model}).
Consider $V(t)$ as an operator in $\KK$ and denote by
$V(m,n),\ m,n\in\Z^2$, its matrix elements in the eigen-basis of $K$.
We have
\be
\label{V}
V(m,n)=\cases{1 & if $m-n\in\{\pm(1,1),\ \pm(1,-1)\}$ ,\cr
0 & otherwise.\cr}
\ee
Concerning the eigen-values of $K$, there is a degeneracy
$$
F(n_1,n_2)=F(n_1,-n_2)=\w n_1+n_2^{\,2}\,.
$$
Let $\LB=\Z (1,1)+\Z (1,-1)$ be a sublattice in $\Z^2$ and denote by
$\PP_0(N)\subset(\Z^2)^{N+1}$ the set of closed paths in $\LB$ of length
$N$ with the base point $\bo$:
$(\io(0),\io(1),\dots,\io(N))\in\PP_0(N)$ iff $\io(0)=\io(N)=\bo$,
$\io(j)\not=\bo$ for
$1\le j\le N-1$, $\io(j)-\io(j-1)\in\{\pm(1,1),\ \pm(1,-1)\}$ for
$1\le j\le N$.
Note that $\PP_0(N)=\emptyset$ for $N$ odd. Clearly,
\be
\label{spec}
\langle QVf,\,\HK^{-\mu_1}\HV\HK^{-\mu_2}\dots\HV\HK^{-\mu_k}QVf\rangle
=\sum_{\io\in\PP_0(k+1)}\,\prod_{j=1}^k F(\io(j))^{-\mu_j} \,.
\ee
The only thing we can claim at this moment is that all $\xi_M$,
$2\le M$, are finite for the sum on the RHS of (\ref{xi1}) is finite.

\section{Diophantine estimates}%

In order to cope with small denominators we need diophantine estimates.
Suppose that we are given two sequences $\psi$ and $E$ such that
$$
\psi: \N\to \,]0,1/2\,]\,,\quad\sum_{k\in\Nin}\psi(k)<\infty,
$$
and
$$
E:\N\to\,]0,+\infty[\,,\quad\inf_{k\in\Nin}E(k)=:d_E>0\,.
$$
Set $F(n):=\w n_1+E(n_2),\ n\in\Z\times\N$, and to a constant $\ga>0$
relate the set
$$
\Omega(\ga):=
\{\w>0;\ \forall n\in\Z\times\N,\ |F(n)|\ge \w\ga\psi(n_2)\}\,.
$$
It is quite standard to show
\proclaim{Lemma 3}.
If $\ga\le d_E/a\le1$ then
$$
|\,]0,a\,]\setminus\Omega(\ga)|\le
\left(16a\,\sum_{k\in\Nin}\psi(k)\right)\,\ga\,.
$$

We can now introduce the set $\Omega$ (depending on $\psi$) of
"non-resonant" frequencies,
$$
\Omega:=\{\w>0;\ \inf_{n\in\Zin\times\Nin}|F(n)|/\psi(n_2)>0\}
=\bigcup_{\ga>0}\Omega(\ga)\,.
$$
As an immediate consequence of Lemma 3 we have
\proclaim{Lemma 4}.
The complement $]0,+\infty[\,\setminus\Omega$ is of zero measure in the
Lebesgue sense.

In the case of our model, $E(k)=k^2$. Extend the definition of $\psi$ by
$\psi(0)=1$ and we define also $F((k,0)):=\w k$. We {\em fix} once for all
$\w\in\Omega$ (and we don't emphasize this fact anymore in the rest of the 
paper).  Then there exists $\ga,\ 0<\ga\le1$, such that
$$
|F(n)|\ge \w\ga\,\psi(|n_2|),\ \forall n\not=\bo\,.
$$

Rather than treating the formal RS series (\ref{RS}) we wish to attack
the fixed-point equation (\ref{fixed}). This means to cope with
expressions involving the operator $\GL$ and hence the numbers
$(F(n)-\la)^{-1}$ -- the eigen-values of $\GL$. The estimate on
$F(n)-\la$ will be governed by a constant $\ro$ and a sequence
$\tps$ of positive reals and we require
$$
\ro\in[\,0,1\,]\quad\mbox{and}\quad\tps(k)\le\psi(k)/2,\
\forall k\in\Z_+\,.
$$
For a given sequence $E$ as above we define a set $\La$ of "good"
parameters $\la$,
\be
\label{La}
\La:=\{\la\in\R;\ \forall n\in\Z\times\N,\ |\FL|\ge\w\ga\,(2|\la|/\w)^\ro\tps(|n_2|)\}\,;
\ee
note that $|\FL|\ge \w/2$ for $n_1\not=0,\ n_2=0$ and $|\la|\le\w/2$.
The following lemma is also easy to prove:
\proclaim{Lemma 5}.
If $0<\delta\le 1/4$ then
$$
|\,[-\dl\w,\dl\w\,]\setminus\La|<2\w(2\dl)^\ro\,
\sum_{k\in\Nin,\ \psi(k)<2\dl}\tps(k)\,.
$$

The standard choice for $\psi$ and $\tps$ is
\be
\label{psi}
\psi(k)=k^{-\sg}/2,\ \tps(k)=k^{-\tau}/4,\ \mbox{with }
1<\sg\le\tau\,.
\ee
In this case we get another intermediate result as a direct consequence
of Lemma 5.
\proclaim{Lemma 6}.
If $\tau>1+\sg(1-\ro)$ then 0 is a point of density for the set $\La$,
i.e.,
$$
\lim_{\dl\downarrow0}{1\over 2\dl\w}\,|[\,-\dl\w,\dl\w\,]\cap\La|=1\,.
$$

\nid Suppose that the sequence $E$ obeys the gap condition (\ref{gap})
with $\al>0$. A possible choice of the constants $\sg,\ \tau$ and $\ro$
which suits the assumption of Lemma 6 is
$$
\tau=1+\al,\ 1<\sg<1+\al,\ \mbox{and } \ro=1/\sg\,.
$$
In our model we have effectively $\al=1$ and so we choose
\be
\label{choice}
\tau=2,\ 1<\sg<2,\ \mbox{and } \ro=1/\sg\in\ ]1/2,1[\,.
\ee

Let us now derive some consequences of the above diophantine estimates in
combination with the gap condition (\ref{gap}). Suppose again that the
spectrum of $H$ is pure point and equals $\{E(k)\}_{k\in\Zin_+}$,
$E(0)=0$, and that $E$ obeys the gap condition (\ref{gap}). It is quite
useful to observe that
another inequality follows straightforwardly from (\ref{gap}),
\be
\label{gap2}
\vert E(j)-E(k)\vert \ge
{C_E\over\alpha+1}\vert j-k\vert\max\{j^\alpha,k^\alpha\},
\quad \forall j,k\in\Z_+\,.
\ee
We shall denote by $P_n$, $n\in\Z\times\Z_+$ (or $\Z\times\Z$ in our
model), the eigen-projectors of $K$ corresponding to the eigen-values
$F(n)$; we have $P\equiv P_{\bo}$ with $F(\bo)=0$. We set also
$Q_n:=1-P_n$.

Another important observation coming from the gap condition is that
those eigen-states $P_n$ which can potentially contribute by small
denominators are distributed rather rarely in the half-plane $n_2\ge 0$.
Let $\SS$ designate the set of "critical" indices defined by:
\be
\label{S}
n\in\SS\ \mbox{iff }\ F(n)\in\ ]-\w/2,\w/2\,]\setminus\{0\}\,.
\ee
Clearly, to each $n_2\in\N$ there exists exactly one $n_1\in\Z$
(necessarily $n_1\le 0$) such that $n\in\SS$; $(n_1,0)\not\in\SS$ for
all $n_1\not=0$, and we treat $n=\bo$ separately since it corresponds to
the eigen-state $P$ to be perturbed. Furthermore, if $m,n\in\SS$ and
$m_2\le n_2$ then $|m_1|\le|n_1|$. Roughly speaking, the indices from
the set $\SS$ are situated closely to the curve $n_1=-E(n_2)/\w$. We set
$\PS:=\sum_{n\in\SS}P_n$, $\QS:=Q-\PS$. Evidently,
$\|\Ga_0\QS\|\le 2/\w$.

Let us introduce a function defined on $\SS$,
\be
\label{L}
L(n):=\min\{|n_2|,\ d(n) \}\,,
\ee
with $\pr_1$ being the projection onto the first coordinate axis, and:
$$
d(n) := \dn = \min_{n'\in\SS,\ |n'_2-n_2|=1} |n'_1-n_1| \leq 
\dist(n_1,\pr_1(\SS)\setminus\{n_1\})\,.
$$
\proclaim{Lemma 7}.
Assume that the function $\tps$ occurring in the definition (\ref{La}) of
the set $\La$ satisfies
$$
\sup_{k\in\Nin}\ k^{-\min\{1,\al\} }\,|\log\tps(k)|<\infty \,.
$$
Then there exists a constant $C_1>1$ such that
$$
|\FL|\ge\lw^\ro\,C_1^{\,-L(n)}\quad\mbox{for }\forall n\in\SS,\
\forall\la\in\La\,.
$$

\nid{\em Proof}. It is sufficient to find $C_1$ so that
$$
\w\ga\,\tps(n_2)\ge\max\{C_1^{\,-n_2},\ C_1^{\,-d(n)}\},\
$$
holds for all $n\in\SS$. Observe that for any couple $m,n\in\SS$,
$m\not=n$, we have $m_2\not=n_2$ and
$$
\w|n_1-m_1|\ge|E(n_2)-E(m_2)|-|\FL|-|F(m)-\la| \,,
$$
and consequently, in virtue of (\ref{gap2}) and the definition (\ref{S})
of $\SS$,
\be
\label{dn}
d(n)\ge(C_E/(\al+1))\,|n_2|^\al-\w\,.
\ee
The rest of the proof is evident. \QED

We are going to verify one more estimate related to the function $L(n)$
defined in (\ref{L}). To this end we shall need
\proclaim{Lemma 8}.
Let $\DL_0,\DL_1,\dots,\DL_\ell$ be a family of positive numbers. Then it
holds
$$
\left|\frac{1}{\DL_1+\DL_2+\dots+\DL_\ell}-{1\over \ell\DL_0}\right|\le
\max_{1\le k\le\ell}\,\left|{1\over\DL_k}-{1\over\DL_{k-1}}\right| \,.
$$

\nid{\em Proof}. The proof follows immediately from the identity
\bea
\frac{1}{\DL_1+\DL_2+\dots+\DL_\ell}-{1\over \ell\DL_0} &=&
{1 \over l}\left[
\left({1\over\DL_1}-{1\over\DL_0}\right)(\DL_1+\dots+\DL_\ell)
\right. \cr
&& +
\left({1\over\DL_2}-{1\over\DL_1}\right)(\DL_2+\dots+\DL_\ell)\cr
&& +\dots\dots \cr
&& \left.
+\left({1\over\DL_\ell}-{1\over\DL_{\ell-1}}\right)\DL_\ell
\right] \,
{1\over \DL_1+\dots+\DL_\ell} \,.\ \QED \nonumber
\eea

Let us define
$$
\DL E(k):=E(k+1)-E(k),\ k\in\Z_+\,,
$$ and suppose that $E$ still satisfies the gap condition (\ref{gap}),
$E(0)=0$. Concerning the function $\tps$ we assume that it is decreasing
and
\be
\label{assum1}
\sup_{k\in\Nin}\tps(k/2)/\tps(k)=:C_\psi<\infty\,.
\ee
The following lemma contains a condition relating the sequences $\DL E$
and $\tps$.
\proclaim{Lemma 9}.
Assume that
\be
\label{assum2}
\sup_{k\in\Zin_+}{1\over\tps(k)}\,
\left|{1\over\DL E(k+1)}-{1\over\DL E(k)}\right|
=:C_\DL<\infty\,.
\ee
Then there exists a constant $C_2>0$ such that for each $n\in\SS$
verifying
\be
\label{assum3}
\min\{\DL E(n_2),\ \DL E(n_2-1)\}\ge4\w\,,
\ee
for all $m\in\Z\times\N,\ m\not=n$, from the neighbourhood
\be
\label{lL}
2\max\{|n_1-m_1|,|n_2-m_2|\}\le L(n)\,,
\ee
and for all $\la\in\La\cap[\,-\w/3,\w/3\,]$ it holds true that
$$
\left|{1\over F(m)-\la}+{1\over F(m')-\la}\right|\le
C_2\,\lw^{-\ro}\,|\FL|\,,
$$
where $m'=2n-m$.

\nid{\em Proof}. The assumptions have some obvious consequences. First,
$$
2|n_1-m_1|\le\dn,\ \mbox{and } m\not=n,
$$
implies that $m\not\in\SS$. Thus one finds that
$$
|F(m)-\la|\ge\left({1\over2}-{1\over3}\right)\w={1\over6}\,\w\,.
$$
Obviously, (\ref{lL}) also implies that $n_2/2\le m_2\le 3n_2/2$.

Furthermore, we have
\be
\label{expr1}
|F(m)-\la|\ge|E(m_2)-E(n_2)|/6\,.
\ee
Indeed, if $m_2\not=n_2$ then
$$
|F(m)-\la|\ge|E(m_2)-E(n_2)|\,\left( 1-
\frac{\w|m_1-n_1|+|F(n)|+|\la|}{|E(m_2)-E(n_2)|} \right) \,.
$$
Let $n'\in\SS$ be such that $|n'_2-n_2|=1$ and
$\sgn(n'_2-n_2)=\sgn(m_2-n_2)$. Then $\dn\le|n_1-n'_1|$ and, owing to
(\ref{lL}),
\bea
2\w|n_1-m_1| &\le& \w|n_1-n'_1|=|E(n'_2)-E(n_2)+F(n)-F(n')| \cr
&\le& |E(m_2)-E(n_2)|+\left({\w\over2}+{\w\over2}\right)\,. \nonumber
\eea
Note that ($m_2\not=n_2$)
$$
|E(m_2)-E(n_2)|\ge\min\{\DL E(n_2),\ \DL E(n_2-1)|\ge 4\w\,.
$$
Altogether this means that
$$
\frac{\w|m_1-n_1|+|F(n)|+|\la|}{|E(m_2)-E(n_2)|}\le
{1\over2}+\left({\w\over 2}+{\w\over 2}+{\w\over 3}\right)\,
{1\over 4\w}={5\over6}
$$
and (\ref{expr1}) follows. All the above estimates are also valid for
$m'$.

Write now
$$
{1\over F(m)-\la}+{1\over F(m')-\la}=
\frac{2(\FL)+E(m_2)+E(m'_2)-2E(n_2)}{(F(m)-\la)(F(m')-\la)}\,.
$$
Now to finish the proof, it suffices to study the case
$m_2-n_2=n_2-m'_2\not=0$. From (\ref{expr1}) one finds that
\bea
\label{expr2}
6^{-2}
\left|\frac{E(m_2)+E(m'_2)-2E(n_2)}{(F(m)-\la)(F(m')-\la)}\right|
&\le& \!\left|{1\over E(m_2)-E(n_2)}+{1\over E(m'_2)-E(n_2)}\right|
\cr &\le&
\left|{1\over E(m_2)-E(n_2)}-{1\over (m_2-n_2)\DL E(n_2)}\right|
\cr && \!\!\!\!+
\left|{1\over E(m'_2)-E(n_2)}-{1\over (m'_2-n_2)\DL E(n_2)}\right|
\eea
Combining Lemma 8, the monotone behaviour of $\tps$, and the
assumption (\ref{assum2}) we get
\bea
\left|{1\over E(j+\ell)-E(j)}-{1\over \ell\,\DL E(j)}\right|
&\le& C_\DL\,\tps(j), \cr
\left|{1\over E(j)-E(j-\ell)}-{1\over \ell\,\DL E(j)}\right|
&\le& C_\DL\,\tps(j-\ell) \,. \nonumber
\eea
Thus we can estimate from above the RHS of (\ref{expr2}) by (c.f.
(\ref{assum1}))
\bea
2C_\DL\,\tps(\min\{m_2,m'_2\}) &\le& 2C_\DL\,\tps(n_2/2) \le
2C_\DL C_\psi\,\tps(n_2) \cr
&\le& (2C_\DL C_\psi/\w\ga)\,\lw^{-\ro}\,|\FL|\,.
\nonumber
\eea
This completes the proof. \QED

Finally note that, with the choice of $\tps$ (\ref{psi}) and for
$E(k)=k^2$, the assumptions of both Lemma 7 and Lemma 8 are satisfied.
Thus these two lemmas are applicable to our example provided the choices
(\ref{psi}) and (\ref{choice}) have been made.

\section{Solution of the fixed-point equation}%

We wish to justify the power series
\be
\label{gbl2}
\gbl=\sum_{k=0}^\infty (-\bt)^{k+1}\,(\GL\HV)^k\,
\GL QVf
\ee
as a solution to the vector equation (\ref{basic}). We start from an
estimate whose proof relies heavily on the very special features of our
model. This doesn't concern the spectrum of $H$ (the gap condition
(\ref{gap}) would be sufficient) but what is really special is the form
of the potential (\ref{V}). For each $m\in\Z^2$ there exist exactly four
indices $n\in\Z^2$ such that $V_{mn}\not=0$. This fact makes it possible
to use some elementary combinatorics in order to treat the summands in
(\ref{gbl2}). The heart of the proof is a sort of compensation based on
Lemma 9. This method of compensations is inspired by the pioneer work of
\ct {Eliasson (1988)}.

Recall the definition of the lattice $\LB$ (Sec.2) and denote by
$\PP(N)\subset(\Z^2)^{N+1}$ the set of (unclosed) paths in $\LB$ of
length $N$ with the initial vertex $\bo$:
$(\io(0),\io(1),\dots,\io(N))
\in\PP(N)$ iff $\io(0)=\bo,\
\io(j)\not=\bo$ for $1\le j\le N$, and
$\io(j)-\io(j-1)\in\{\pm(1,1),\pm(1,-1)\}$ for $1\le j\le N$.
Clearly, $|\PP(N)|\le 4^N$. For $M\in\N$ one can write
\be
\label{path}
(\GL\HV)^{M-1}\GL QVP=\sum_{\io\in\PP(M)}
\left(\prod_{j=1}^M{1\over F(\io(j))-\la}\right)\,P_{\io(M)}\,.
\ee
\proclaim{Lemma 10}.
In the case of the model (\ref{model}) and assuming that the choices
(\ref{psi}) and (\ref{choice}) have been made, there exists a constant
$\HC>0$ such that
$$
\|\GL QVf\|\le\HC,\ \ \|(\GL\HV)^{M-1}\GL QVf\|\le\lww^\ro
\left(\lww^{-\ro/2}\HC\right)^M
$$
holds true for $\forall M\in\N,\ M\ge2$, and
$\forall\lambda\in\La\cap[\,-\w/3,\w/3\,],\ \la\not=0$.

\nid{\em Remark}. Note the type of the estimate: we are able to estimate
the vector
\newline
$(\GL\HV)^{M-1}\GL QVf$ but not directly the operator
$(\GL\HV)^M$.

\nid{\em Proof}. We start from restricting the set $\SS$ of critical
indices to a subset $\SS'=\{n\in\SS;\ |n_2|>b\}$. The bound $b\in\N$ is
required to obey the conditions: \newline
$\bullet$ $b\ge3$, \newline
$\bullet$ $4\w\le\min\{\DL(k),\DL(k-1)\}$ for $\forall k>b$, \newline
$\bullet$ $L(n)\ge2$ for $\forall n\in\SS,\ |n_2|>b$. \newline
The second requirement is dictated by the assumption (\ref{assum3}) of
Lemma 9 and the third one is possible since from the estimate (\ref{dn})
follows that
$$
\lim_{n\in\SS,\ |n_2|\to\infty} L(n)=+\infty\,.
$$

Clearly, since $|\FL|\ge\w/6$ for $n\not\in\SS,\ |\la|\le\w/3$, there
exists a constant $C_3>0$ such that
$$
|\FL|\ge C_3\quad\mbox{for } \forall n\not\in\SS',\
\forall\la\in\La\cap[\,-\w/3,\w/3\,]\,.
$$

Without loss of generality we can restrict ourselves to $M\ge2$. For
each $\io\in\PP(M)$ the vertices from $\SS'$ split the path into
segments. Consider such a segment of length $\ell$,
$(\io(j),\io(j+1),\dots,\io(j+\ell))$, with $\io(j+\ell)\in\SS'$, and
also $\io(j)\in\SS'$ provided $j\not=0$, and $\io(j+s)\not\in\SS'$ for
$1\le s\le\ell-1$. However, in order not to count it twice, we don't
relate to the segment the contribution from the vertex $\io(j)$.

We distinguish two cases. If $\ell\ge L(\io(j+\ell))$ then Lemma 7
implies
\be
\label{expr3}
\left|\prod_{s=j+1}^{j+\ell}{1\over F(\io(s))-\la}\right|\le
\left({1\over C_3}\right)^{\ell-1}\,\lww^{-\ro}\,C_1^{\,\ell} \,.
\ee

Consider now the case $\ell<L(\io(j+\ell))$. The possibility $j=0$ is
excluded since this would imply $\ell<|\io(\ell)_2|\le\ell$. Thus
$\io(j),\io(j+\ell)\in\SS'$ and necessarily $\io(j)=\io(j+\ell)$ as
follows from
$$
|\io(j+\ell)_1-\io(j)_1|\le\ell<
\dist(\io(j+\ell)_1,\ \pr_1(\SS)\setminus\{\io(j+\ell)_1\}) \,.
$$
Consequently, $\ell$ is even. We shall call a segment of this type short
loop. To any short loop there exists an opposite short loop
$(\io'(j),\io'(j+1),\dots,\io'(j+\ell)=\io'(j))$ defined by
$\io'(s):=2\io(j)-\io(s),\ j\le s\le j+\ell$; hence the base point is
the same, $\io'(j)=\io(j)$. Now we are approaching the compensation
step. The contribution of two opposite short loops equals
\bea
\label{expr4}
&& \prod_{s=j+1}^{j+\ell}{1\over F(\io(s))-\la}+
\prod_{s=j+1}^{j+\ell}{1\over F(\io'(s))-\la} \cr
&& = {1\over F(\io(j))-\la} \left(
\prod_{s=j+1}^{j+\ell-1}{1\over F(\io(s))-\la}-
\prod_{s=j+1}^{j+\ell-1}{1\over -F(\io'(s))+\la} \right) \,.
\eea
In order to estimate the difference of products on the RHS of
(\ref{expr4}) one can use the identity
\be
\label{tele}
u_1\dots u_N-v_1\dots v_N=\sum_{s=1}^N
u_1\dots u_{s-1}(u_s-v_s)v_{s+1}\dots v_N
\ee
and Lemma 9. This way one arrives at
\be\label{expr5}
|\mbox{expression}(\ref{expr4})|\le (\ell-1)
\left({1\over C_3}\right)^{\ell-2}\,C_2\,\lww^{-\ro}\le
C_2C_3^{\,2}\,\lww^{-\ro}\,\left({2\over C_3}\right)^{\ell}\,.
\ee

In order to treat this type of compensation systematically let us split
$\PP(M)$ into equivalence classes. Two paths are equivalent if and only
if one is obtained from the other by replacing several short loops by
their opposites. Thus a path containing $s$ short loops belongs to a
class with $2^s$ elements. One can write schematically
$$
\sum_{\mbox{\footnotesize all paths}}\ \ \prod_{\mbox{\footnotesize all 
segments}} =\sum_{\mbox{\footnotesize equivalence classes}}\ \ 
\prod_{\mbox{\footnotesize pairs of short loops}}\ \times\ 
\prod_{\mbox{\footnotesize other segments}}
$$
For a path $\io\in\PP(M)$ denote by $N=N(\io)$ the number of vertices
belonging to $\SS'$. Obviously, $N(\io)$ is constant an every equivalence 
class. Relying on the estimates (\ref{expr3}) and
(\ref{expr5}) one concludes readily that there exists a constant $\HC>0$
such that
$$
\left|\sum_{\mbox{\footnotesize equivalence class}}\ \ \prod_{j=1}^M
{1\over F(\io(j))-\la} \right| \le
\lww^{-\ro N}\,\left({\HC\over4}\right)^M\,.
$$
Since $b\ge3$ we have $\io(1),\io(2),\io(3)\not\in\SS'$ and
consequently, as $L(n)\ge2$ for all $n\in\SS'$,
$$
2N(\io)\le M-2\,.
$$
To complete the proof it suffices to estimate from above the number of
equivalence classes simply by $|\PP(M)|\le 4^M$ (c.f. (\ref{path})).
\QED

With the estimate given in Lemma 10, it is quite straightforward to
derive the following existence (but not uniqueness) result.
\proclaim{Lemma 11}.
Under the same assumptions as in Lemma 10, the series (\ref{gbl2}) converges
to a solution $\gbl$ of the equation (\ref{basic}) provided $(\bt,\la)$
belongs to the domain
\be
\label{domain}
\la\in\La\cap[\,-\w/3,\w/3\,],\ |\bt|\le\lw^{\ro/2}/2\HC \,.
\ee
For each $\la\in\La\cap[\,-\w/3,\w/3\,],\ \la\not=0$, the vector-valued
function $\gbl$ is analytic in $\bt$ on the corresponding neighbourhood
of 0 and
\be
\label{estimate}
\|\gbl+\bt\GL QVf\|\le 2\HC^2\bt^2\,.
\ee

Now we can give a precise meaning to the RHS of the fixed-point equation
(\ref{fixed}). For $(\bt,\la)$ from the domain (\ref{domain}),
\bea
\label{G}
\Gbl:=\bt\,\langle QVf,\,\gbl\rangle=\sum_{k=1}^\infty\bt^{2k}\,
G_{2k}(\la)\,, \cr
\mbox{where }\ G_{2k}(\la):=-
\langle QVf,\,(\GL\HV)^{2k-2}\GL QVf\rangle\,.
\eea
In our particular example we have
$G_{2k+1}(\la)=0$ for $k\ge1$ but generally
this need not be the case. As a consequence of Lemma 10 we get
\be
\label{estimate2}
|G_{2k}(\la)|\le\|V\|\,\lww^\ro\,
\left(\lww^{-\ro/2}\,\HC\right)^{2k-1}\,.
\ee
Particularly for our model ($E(1)=1$),
$$
G_2(\la)=-\langle QVf,\,\GL QVf\rangle=
\frac{4(E(1)-\la)}{\w^2-(E(1)-\la)^2}\,,
$$
and $G_2(0)\not=0$.

We shall impose a stricter bound on $\la$, $|\la|\le\la_\star$,
where $0<\las\le\w/3$, and we require that $\las$ is sufficiently small
so that \newline
$\bullet$ $|G_2(\la)-G_2(0)|\le |G_2(0)|/2$, \newline
$\bullet$ $(2\las/\w)^{1-\ro}\le |G_2(0)|/(8\w\HC^2)$, \newline
$\bullet$ $\las^{1/2}\le |G_2(0)|^{3/2}/(16\,\|V\|\,\HC^2)$,  \newline
$\bullet$ $(2\las/\w)^{\ro/2}\le |G_2(0)|/(2\,\|V\|\,\HC)$.  \newline
Recall that $1/2<\ro<1$ (c.f. (\ref{choice})). Set
$$
B(\la):=2\,(|\la|/|G_2(0)|)^{1/2}\,.
$$
The first requirement implies $|G_2(\la)|\ge|G_2(0)|/2$ and
$\sgn\,G_2(\la)=\sgn\,G_2(0)$. Owing to the second requirement we have
$$
|\la|\le\las \Longrightarrow \BL\le\lw^{\ro/2}/2\HC
$$
and so $\la\in\La\cap[\,-\las,\las\,],\ |\bt|\le\BL$ determines a
subdomain of (\ref{domain}). From the third requirement follows that
\be
\label{third}
|\la|\le\las\Longrightarrow 2\|V\|\,\HC^2\BL^3\le|\la|\,.
\ee
Finally, a routine calculation based on the definition (\ref{G}) of $G$,
the estimate (\ref{estimate2}), and the fourth requirement yields the
inequality
\be
\label{diff}
|\pa_\bt\Gbl-2\bt G_2(\la)|<|\bt|\,|G_2(0)|\le 2|\bt|\,|G_2(\la)|,
\ee
valid for $0<|\la|\le\las,\ 0<|\bt|\le\lw^{\ro/2}/2\HC$. Consequently,
\be
\label{mono}
\sgn\,\pa_\bt\Gbl=\sgn\,\bt G_2(\la)=\sgn \bt G_2(0)\,.
\ee
\proclaim{Lemma 12}.
Under the same assumptions as in Lemma 10, for each
$\la\in\La\cap[\,-\las,\las\,]$, $\sgn\,\la=\sgn\,G_2(0)$, there exist
exactly two solutions $\bpl$ to the equation $\la=\Gbl$ in the interval
$[\,-\BL,\BL\,]$, and there is no solution for
$\sgn\,\la=-\sgn\,G_2(0)$. The two solutions are non-zero, differ in
sign, and we choose the convention
$$
-\BL\le\bt_-(\la)<0<\bt_+(\la)\le\BL\,.
$$
Then $\la$ is an eigen-value of the operators $K+\bt_\pm(\la)\,V$.

\nid{\em Remark}. Since, in the case of our model, $\Gbl$ is even in $\bt$
we have consequently $\bt_-(\la)=-\bt_+(\la)$. But, of course, this is
not a general feature.

\nid{\em Proof}. Obviously, $G(0,\la)=0$. Let us show that
$|G(\pm\BL,\la)|\ge |\la|$. From (\ref{estimate}) we obtain
$$
|\Gbl-\bt^2 G_2(\la)|=
|\bt\,\langle QVf,\,\gbl+\bt\GL QVf\rangle|
\le 2\|V\|\,\HC^2|\bt|^3
$$
and, owing to (\ref{third}),
$$
|G(\pm\BL,\la)-\BL^2\,G_2(\la)|\le |\la|\,.
$$
On the other hand,
$$
|\BL^2\,G_2(\la)|\ge 4{|\la|\over |G_2(0)|}\cdot{1\over2}\,|G_2(0)|
=2|\la|\,.
$$
This way we have also verified that
$$
\sgn\,G(\pm\BL,\la)=\sgn\,G_2(\la)=\sgn\,G_2(0)\,.
$$
Now the existence follows from the fact that the function $\Gbl$ is
continuous (even analytic) in $\bt$. The uniqueness is a consequence of
the monotone behaviour (c.f. (\ref{mono})). \QED

\section{Properties of the function $\lb$}%

Inverting the functions $\bt_+(\la)$ and $\bt_-(\la)$ we expect to
obtain the desired function $\lb$ defined respectively on sets $I_+$ and
$I_-$, with $I_\pm\subset\R_\pm$, and we set naturally $\la(0)=0$. Thus
the total domain for $\lb$ is $I=I_-\cup \{0\}\cup I_+$. $\lb$ is
positive (negative), except of $\la(0)=0$, if $G_2(0)$ is positive
(negative). The existence of the inverted function follows from the
monotone behaviour of the original functions $\bpl$.

We shall need
\proclaim{Lemma 13}.
The function $\Gbl$ defined in (\ref{G}) fulfills the equality
$$
G(\bt,\la_2)-G(\bt,\la_1)=-(\la_2-\la_1)\,
\langle g(\bt,\la_2),\,g(\bt,\la_1)\rangle
$$
for all
\be
\label{min}
\la_1,\la_2\in\La\cap [\,-\w/3,\w/3\,],\
|\bt|\le(2\min\{|\la_1|,|\la_2|\}/\w)^{\ro/2}/2\HC\,.
\ee

\nid{\em Proof}. Note that
$\Ga_{\la_2}-\Ga_{\la_1}=(\la_2-\la_1)\,\Ga_{\la_2}\Ga_{\la_1}$ on
$\DD(\Ga_{\la_1})\cap\DD(\Ga_{\la_2})$ and consequently, using
(\ref{tele}),
\bea
&& \langle QVf,\,(\Ga_{\la_2}\HV)^k \Ga_{\la_2}QVf-
(\Ga_{\la_1}\HV)^k \Ga_{\la_1}QVf\rangle \cr
&& =(\la_2-\la_1)\,\sum_{j=0}^k
\langle (\Ga_{\la_2}\HV)^j \Ga_{\la_2}QVf,\,
(\Ga_{\la_1}\HV)^{k-j} \Ga_{\la_1}QVf\rangle \,. \nonumber
\eea
Now the identity can be verified easily with the aid of (\ref{gbl2}).
\QED

From (\ref{estimate}) one deduces that
$\langle g(\bt,\la_2),\,g(\bt,\la_1)\rangle >0$ whenever
$|\la_1|,|\la_2|$ are sufficiently small and $|\bt|$ obeys (\ref{min}).
Thus we find that $\Gbl$ is strictly decreasing in $\la$ for every $\bt$
fixed. The same is true for the function $\Phi(\bt,\la):=\Gbl-\la$.

This is an elementary exercise to verify that the functions $\bpl$ are
strictly monotone provided one uses the equality $\Phi(\bpl,\la)=0$ and
the fact that $\Pbl$ is monotone in $\bt$ (c.f. (\ref{mono})) and
strictly monotone in $\la$. We can formulate our conclusion as follows.
\proclaim{Lemma 14}.
There exists a bound $\la_{\star\star}$, $0<\la_{\star\star}\le\las$,
and a function $\lb$ defined on $I\subset \R$ such that $0\in I$ and
$\la(0)=0$, $\bt_\pm(\lb)=\bt$ for $\forall\bt\in I\cap\R_\pm$, and the
range of both $\lb |I\cap\R_+$ and $\lb |I\cap\R_-$ equals either
$\La\cap[\,0,\la_{\star\star}\,]$ or
$\La\cap[\,-\la_{\star\star},0\,]$ depending on whether $G_2(0)$ is
positive or negative. For $\forall\bt\in I$, $\lb$ is an eigen-value of
the operator $K+\bt V$.

This seems to be a typical feature for the perturbation theory of dense
point spectra that one has to abandon some values of the coupling
constant $\bt$ and to determine the perturbed eigen-value as a function
$\lb$ defined on a domain $I$ possessing "holes". To treat functions of
this type one can refer to the celebrated Whitney Extension Theorem (see \ct 
{Stein  1970}). In fact, its proof in the one-dimensional case is rather
elementary. We  shall need the following very particular version.
\proclaim{Lemma 15}.
Let $\chi$ be a real function defined on a closed subset $Y\subset\R$,
$\chi$ is monotone, and suppose that there exist two constants
$0<A\le B$ such that
$$
A|y_1-y_2|\le|\chi(y_1)-\chi(y_2)|\le B|y_1-y_2|\quad\mbox{for all }
y_1,y_2\in Y \,.
$$
Then there exists an extension $\tch$ defined on $\R$, $\tch|Y=\chi$,
and $\tch$ is again monotone and obeys the same inequalities but this time 
on the  whole line $\R$,
$$
A|y_1-y_2|\le|\tch(y_1)-\tch(y_2)|\le B|y_1-y_2|\quad\mbox{for all }
y_1,y_2\in \R \,.
$$

\nid{\em Proof}. The complement of $Y$ is an open subset of $\R$ and
hence at most countable disjoint union of open intervals. One defines
the function $\tch$ linearly on these intervals requiring it to be
continuous. Provided the interval in question is half-infinite then
$\tch$ is defined again linearly with the slope lying between $A$ and
$B$. The inequalities for $\tch$ defined this way are easy to verify;
for the left one we need that $\chi$ is monotone. \QED

We wish to show that 0 is a point of density for the set $I$. We already
know that this is true for the set $\La$ (Lemma 6). The intermediate
step is given by
\proclaim{Lemma 16}.
Assume that a real function $\vp(x)$, defined on a set
$X\subset[\,0,+\infty[$, is strictly increasing, $\vp(0)=0$
$(\Rightarrow 0\in X$), and the set $Y=\vp(X)$ is closed. Moreover,
suppose that there exist two constants $0<A\le B$ such that
\be
\label{inequal}
A|x_1^{\,2}-x_2^{\,2}|\le |\vp(x_1)-\vp(x_2)|\le
B|x_1^{\,2}-x_2^{\,2}|\quad\mbox{for all } x_1,x_2\in X\,.
\ee
Then it holds
\be
\label{imply}
\lim_{\eta\downarrow0}|Y\cap[\,0,\eta\,]|/\eta=1
\,\Longrightarrow\,
\lim_{\dl\downarrow0}|X\cap[\,0,\dl\,]|/\dl=1 \,.
\ee

\nid{\em Proof}. Apply Lemma 14 to the function
$\chi(y)=(\vp^{-1}(y))^2$ (the corresponding constants are $0 < 1/B \leq 
1/A$). The extension $\tch$ is again strictly
increasing, $\tch(y)>0$ for $y>0$, and $\tch(\R_+)=\R_+$. Define $\tvp$
on $\R_+$ by $\tvp(x)=y$ iff $x^2=\tch(y)$, i.e.,
$\tvp$ is the inverse of
$(\tch|\R_+)^{1/2}$. Clearly, the function $\tvp$ is an extension of
$\vp$, $\tvp|X=\vp$, it is again strictly increasing, and the
inequalities (\ref{inequal}) hold for $\tvp$ on the whole positive
half-line. Consequently, $\tvp$ is absolutely continuous on every
bounded interval, $\tvp'$ exists almost everywhere, and it holds
$$
\tvp(x)\le B\,x^2\ \mbox{and}\ 2Ax\le \tvp'(x)\quad
\mbox{for (almost) all } x\ge 0\,.
$$

Denote by $X^c$ and $Y^c$ the complements of $X$ and $Y$ in
$[\,0,+\infty[$, respectively. The implication (\ref{imply}) is
equivalent to
\be
\label{imply2}
\lim_{\eta\downarrow0}|Y^c\cap[\,0,\eta\,]|/\eta=0
\,\Longrightarrow\,
\lim_{\dl\downarrow0}|X^c\cap[\,0,\dl\,]|/\dl=0 \,.
\ee
Choose $p$, $1<p<2$, and let $q$ be the adjoint exponent,
$p^{-1}+q^{-1}=1$. We shall verify the inequality
\be
\label{expr6}
\dl^{-1}\,|X^c\cap[\,0,\dl\,]|\le {B\over 2A}\,
\left(1-{p\over 2}\right)^{-1/p}\, \left(
\tvp(\dl)^{-1}\,|Y^c\cap[\,0,\tvp(\dl)\,]| \right)^{1/q} \,.
\ee
It is clear that (\ref{imply2}) is a consequence of (\ref{expr6}). We
have
$$
|X^c\cap[\,0,\dl\,]|=\int_{Y^c\cap[\,0,\tvp(\dl)\,]}\,
{dy\over \tvp'(\tvp^{-1}(y))} \le
{\sqrt{B}\over 2A}\,
\int_{Y^c\cap[\,0,\tvp(\dl)\,]}\,y^{-1/2}\,dy
$$
since $\tvp'(\tvp^{-1}(y))\ge 2A\,\tvp^{-1}(y)\ge 2A\,(y/B)^{1/2}$.
H\"older Inequality then gives
$$
\int_{Y^c\cap[\,0,\tvp(\dl)\,]}\,y^{-1/2}\,dy\le
\left(\int_0^{\tvp(\dl)}y^{-p/2}\,dy\right)^{1/p}\,
\left(\int_{Y^c\cap[\,0,\tvp(\dl)\,]}\,dy\right)^{1/q}
$$
and (\ref{expr6}) follows immediately. \QED

Observe that the property (2) given in Proposition 1 is equivalent to
$$
\lim_{\dl\downarrow0}|I\cap[\,0,\dl\,]|/\dl=1\quad\mbox{and}\quad
\lim_{\dl\downarrow0}|I\cap[\,-\dl,0\,]|/\dl=1\,.
$$
Thus we can treat the right and the left neighbourhood of 0 separately.
We can now apply Lemma 16 to the function $\lb$ instead of $\vp(x)$ and
to the sets $I_+\cup\{0\}$ and $I_-\cup\{0\}$ instead of $X$. Observe
from the definition (\ref{La}) that $\La$ is closed. Let us show that
the condition (\ref{inequal}) is fulfilled as well. Assume that
$\bt_1,\bt_2\in I,\ |\bt_1|<|\bt_2|$. Then
$(\bt_1,\la(\bt_1)),\ (\bt_2,\la(\bt_2))$ and $(\bt_1,\la(\bt_2))$
belong to the domain of $G$. Write
$$
\la(\bt_1)-\la(\bt_2)=G(\bt_1,\la(\bt_1))-G(\bt_1,\la(\bt_2))
+G(\bt_1,\la(\bt_2)) -G(\bt_2,\la(\bt_2))
$$
and use Lemma 13 to get
$$
\la(\bt_1)-\la(\bt_2)=
\big( G(\bt_1,\la(\bt_2))-G(\bt_2,\la(\bt_2))\big)/
(1+\langle g(\bt_1,\la(\bt_1)),\,g(\bt_1,\la(\bt_2))\rangle)\,.
$$
Deduce from (\ref{estimate}) that
$$
0< \langle g(\bt_1,\la(\bt_1)),\,g(\bt_1,\la(\bt_2))\rangle
= O(|\bt_2|^2),\ \mbox{as } |\bt_1|\le|\bt_2|\to 0\,,
$$
and note that (\ref{diff}) can be rewritten as
$$
|\pa_{\bt^2}\Gbl-G_2(\la)|\le |G_2(0)|/2\,.
$$
One readily concludes that there exist constants $0<A\le B$ and a bound
$\bt_\star>0$ such that
$$
A|\bt_1^{\,2}-\bt_2^{\,2}|\le |\la(\bt_1)-\la(\bt_2)|\le
B|\bt_1^{\,2}-\bt_2^{\,2}|\quad\mbox{for all }
\bt_1,\bt_2\in I\cap[\,-\bt_\star,\bt_\star\,]\,.
$$
\proclaim{Lemma 17}.
0 is a point of density for the set $I$.

Now we can approach the problem of the asymptotic series. Consider first
the following situation. Let $\{ H_k\}_{k=0}^\infty$ be a sequence of
complex meromorphic functions such that $H_0'(0)\not=0$ and 0 is a
regular point for all of them. Then
$$
\Phi(x,y):=\sum_{k=0}^\infty x^k\,H_k(y)\in\C[[x,y]]
$$
is well defined as a formal power series in $x$ and $y$. Denote by
$\vf(x)\in\C[[x]]$ the solution to the problem
$$
\vf(0)=0\quad\mbox{and}\quad \Phi(x,\vf(x))=0\,,
$$
which exists and is unique in the class of formal power series. Set
$$
\RR_\Phi:=\C\setminus\bigcup_{k=0}^\infty
\{\mbox{the poles of the function }H_k\}
$$
and let $R(y)$ be the radius of convergence of the series $\Phi(x,y)$ in
the variable $x$, with $y\in\RR_\Phi$ being fixed.
\proclaim{Lemma 18}.
Let $\vp$ be a complex function defined on $X\subset\C$ and assume
that: \newline
(1) $0\in X$ is an accumulation point of $X$, \newline
(2) $\forall x\in X,\ |x|<R(\vp(x))$ (and so the value $\Phi(x,\vp(x))$
is well defined), \newline
(3) $\vp$ solves the problem
$$
\vp(0)=0\quad \mbox{and}\quad \Phi(x,\vp(x))=0\quad\mbox{for }
\forall x\in X\,,
$$
\newline
(4) there exists $\mu>0$ such that
\bea
&& \Phi_N(x,\vp(x))=O(|x|^{\mu(N+1)})\quad\mbox{for }
\forall N\in\Z_+,\ \mbox{where } \cr
&& \Phi_N(x,y):=\sum_{k=0}^N x^k\,H_k(y)\,. \nonumber
\eea
Then $\vf(x)$ is an asymptotic series for $\vp(x)$.

\nid{\em Proof}. Denote by $\vf_M$ the truncation of $\vf$ (thus $\vf_M$
is a polynomial of degree at most $M$ and $\vf(x)-\vf_M(x)\in
x^{M+1}\,\C[[x]]$). We have to show that
$$
\vp(x)-\vf_M(x)=O(|x|^{M+1}),\ \forall M\in\Z_+\,.
$$
Denote by $\vn(x)$ the unique solution to the problem
$$
\vn(0)=0\quad\mbox{and}\quad \Phi_N(x,\vn(x))=0\,,
$$
in the class of germs of holomorphic functions at $x=0$. Clearly,
$$
\vf_M(x)=\vn_M(x)\quad\mbox{if } N\ge M\,.
$$

Note that the requirement (4), with $N=0$, means that
$H_0(\vp(x))=O(|x|^\mu)$. Since $H_0'(0)\not=0$ we find that
$\lim_{x\to 0}\vp(x)=0$. Obviously, it also holds that
$\lim_{x\to 0}\vn(x)
=0$.
Observe that $\pa_y\Phi_N(0,0)=H_0'(0)\not=0$.
Consequently, for any $n\in\Z_+$, there exist positive constants  $c_N,\
\dl_N$ such  that
$$
|\Phi_N(x,\vp(x))-\Phi_N(x,\vn(x))|\ge c_N\,|\vp(x)-\vn(x)|\quad
\mbox{for }\forall x\in X,\ |x|\le\dl_N\,.
$$

Fix $M\in\Z_+$ and choose $N\in\Z_+$ such that $N\ge M$ and
$\mu(N+1)\ge M+1$. Write
$$
\vp(x)-\vf_M(x)=\vp(x)-\vn(x)+\vn(x)-\vn_M(x)=\vp(x)-\vn(x)+
O(|x|^{M+1})\,.
$$
On the other hand,
$$
c_N\,|\vp(x)-\vn(x)|\le |\Phi_N(x,\vp(x))-\Phi_N(x,\vn(x))| =
|\Phi_N(x,\vp(x))|=O(|x|^{\mu(N+1)})\,.
$$
We conclude that $\vp(x)-\vf_M(x)=O(|x|^{M+1})$, as required. \QED

Lemma 17 is directly applicable to the function
$\Phi(\bt,\la):=\Gbl-\la$ and to our solution $\lb$.
\proclaim{Lemma 19}.
The formal power series $\sum_{M=0}^\infty \xi_M\,\bt^M$, with $\xi_M$
given in (\ref{xi1}) and (\ref{spec}), is an asymptotic series for the
function $\lb$ defined on $I$.

Let us summarize that Lemma 14, Lemma 17 and Lemma 19 verify jointly the
existence and the properties of the function $\lb$ and thus the proof of
Proposition 1 has been completed.


\vskip 24pt
\nid{\bf Acknowledgements.}\
P. S. wishes to express his gratitude to his hosts at  Centre de
Physique Th\'eorique in Marseille and at Universit\'e de Toulon et du Var.
The support from Grant No. 201/94/0708 of Czech GA is also gratefully
acknowledged.

\pagebreak

\end{document}